REPORT

# Monitoring the daily variation of Sun-Earth magnetic fields using galactic cosmic rays

The LHAASO Collaboration[1,*]

*Correspondence: nanyc@ihep.ac.cn (Y.N.); chensz@ihep.ac.cn (S.C.); fengcf@sdu.edu.cn (C.F.)

## GRAPHICAL ABSTRACT

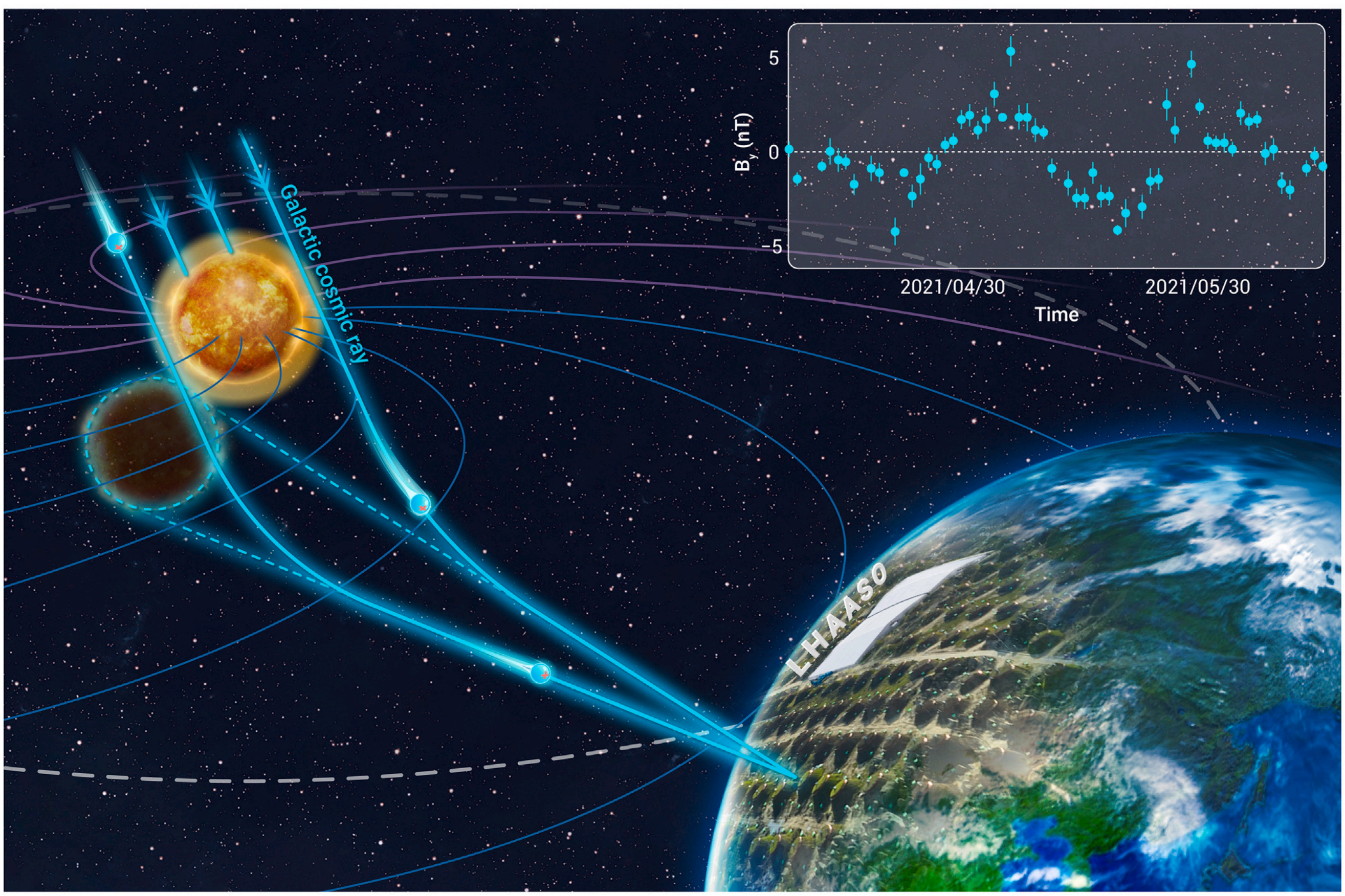

## PUBLIC SUMMARY

- Daily Sun's shadows on very high-energy galactic cosmic rays were observed for the first time using LHAASO.
- This work therefore proposes a novel method for monitoring the Sun-Earth IMF.
- Compared with the near-Earth spacecraft, the Sun's shadow can provide 3.3-day earlier predictions for the IMF.
- The timing advance significantly deviated from the predictions of current IMF models.
- These findings may provide valuable insights into the IMF structure, thus improving space weather research.

# Monitoring the daily variation of Sun-Earth magnetic fields using galactic cosmic rays

**The LHAASO Collaboration**[1,*]

[1]Further details can be found in the supplemental information

*Correspondence: nanyc@ihep.ac.cn (Y.N.); chensz@ihep.ac.cn (S.C.); fengcf@sdu.edu.cn (C.F.)

The interplanetary magnetic field (IMF) between the Sun and Earth is an extension of the solar magnetic field carried by the solar wind into interplanetary space. Monitoring variations in the IMF upstream of the Earth would provide very important information for the prediction of space weather effects, such as effects of solar storms and the solar wind, on human activity. In this study, the IMF between the Sun and Earth was measured daily for the first time using a cosmic-ray observatory. Cosmic rays mainly consist of charged particles that are deflected as they pass through a magnetic field. Therefore, the cosmic-ray Sun shadow, caused by high-energy charged cosmic rays blocked by the Sun and deflected by the magnetic field, can be used to explore the transverse IMF between the Sun and Earth. By employing the powerful kilometer-square array at the Large High Altitude Air Shower Observatory, the cosmic-ray Sun shadows were observed daily with high significance for the first time. The displacement of the Sun shadow measured in 2021 correlates well with the transverse IMF component measured *in situ* by spacecraft near the Earth, with a time lag of $3.31 \pm 0.12$ days. The displacement of the Sun shadow was also simulated using Parker's classic IMF model, yielding a time lag of $2.06 \pm 0.04$ days. This deviation may provide valuable insights into the magnetic field structure, which can improve space weather research.

## INTRODUCTION

The Sun, our nearest star, is the main source of energy for living organisms on Earth, and its activity continually affects our planet's environment. As human science and technology advance, along with the increasing use of electronic and space equipment, the impact of solar activity on human activity has steadily grown. Therefore, monitoring solar activity and forecasting space weather are important areas of scientific research.

Solar magnetic fields play a vital role in understanding diverse solar activities. Since the first measurement of the solar magnetic field in 1908 using the Zeeman effect,[1] the photospheric magnetic field at the Sun's surface has been continuously monitored by both space- and ground-based observatories.[2] The coronal magnetic field (CMF) lies above the photosphere, and direct measurement remains challenging, despite recent attempts using magnetoseismology.[3] The CMF is carried into interplanetary space by the solar wind, forming the interplanetary magnetic field (IMF).[4] The IMF provides valuable information for studying the CMF and is critical for understanding space weather and improving forecasting accuracy.[5,6] Since the discovery of the solar wind in 1962, the IMF has been monitored *in situ* by a series of spacecraft, with considerable monitoring performed from the $L_1$ Lagrange point of gravitational balance between the Sun and Earth.[7]

Although the ongoing Parker Solar Probe mission can fly from the IMF to the CMF,[8] at locations other than the Sun's surface and $L_1$ point, it remains challenging to continuously monitor the magnetic field in the vast space between the Sun and Earth. Currently, the distribution of this magnetic field relies on theoretical models that extrapolate the photospheric magnetic field to the CMF (e.g., the classical potential field source surface [PFSS] model[9,10]) and extend the outermost CMF to the IMF (e.g., the classical and widely used Parker model[4]). The distributions of the CMF and IMF can also be simulated using data-driven models.[11]

Very-high-energy galactic cosmic rays, consisting mainly of positively charged particles moving near the speed of light, can travel from the Sun to the Earth within approximately 8 min. Their trajectories are affected by the magnetic field along the Sun-Earth line. Therefore, when the cosmic-ray Sun shadow was detected for the first time, it was proposed that the magnetic field between the Sun and Earth could be studied through measurements of the Sun shadow.[12] It was noted that as observational sensitivity increases, the cosmic-ray deficit ratio of the Sun shadow was observed or proved to have a relationship with the solar magnetic field, including the photospheric magnetic field,[13-15] CMF and its annual variations,[16-19] and even coronal mass ejections (CMEs),[20] and it was applied to diagnose different CMF models.[16,17]

In addition, studies have observed[21-25] and proved[23-26] that the displacement of the Sun shadow is also related to the IMF. Based on the annual timescale displacement of the Sun shadow, the ARGO-YBJ collaboration measured the structure and strength of the mean IMF near the sunspot minimum in Solar Cycle 24 for the first time.[24] They also proposed the possibility of using Sun shadow measurements for space weather forecasting, given the time advantage this method offers over spacecraft at $L_1$. However, this approach requires cosmic-ray arrays with enough sensitivity to measure shadows daily.[24] In addition, the ASγ collaboration used annual Sun shadow data to correct the average strength of the Parker model for the IMF near total Solar Cycle 23.[25]

Owing to the limited observational sensitivity of the Sun shadows, researchers have not yet been able to study finer structures in the magnetic field, particularly during short periods.

## RESULTS AND DISCUSSION

The Large High Altitude Air Shower Observatory (LHAASO) is a composite ground-based cosmic-ray detection facility located at 100.01°E, 29.35°N, and an altitude of 4,410 m above sea level in Sichuan, China.[27,28] The 1.3 kilometer square array (KM2A) is one of the main ground-based arrays in LHAASO, which has a detection area of 1 to 2 orders of magnitude larger than those of the ARGO-YBJ and ASγ experiments. KM2A has been operating with a nearly full-duty cycle since the beginning of 2020. In this study, only the data recorded in 2021 near the sunspot minimum in Solar Cycle 25 were used. Owing to the location of the array near the Tropic of Cancer, the Sun shadow could be observed significantly for 196 days from March 21 to October 2, with ∼7 h observation on each day. During this period, the pointing accuracy, angular resolution, and energy of the cosmic rays measured by the array were stable according to the observation of the cosmic-ray Moon shadow.[29] For monitoring the IMF, we selected events with 26–251 fired electromagnetic particle detectors (EDs) in KM2A. The corresponding median energy was ∼40 TeV, and the angular resolution was 0.5°.

The Sun shadow was observed along the longitude and latitude in the geocentric solar ecliptic (GSE) coordinate system. Figure 1A depicts the Sun shadows observed during the 196 days of data collection, with a very high significance exceeding 100σ. Figure 1B displays the Sun shadow obtained from a single day of data, which also demonstrates good significance exceeding 9σ. The daily Sun shadows are clearly deflected away from the direction of the Sun by the magnetic field. The angular distance from the center of the shadow to the Sun is defined as the daily displacement of the shadow. Details regarding the Sun shadow analysis are presented in the materials and methods.

### Daily IMF-$B_y$ measurement using Sun shadow

Cosmic rays traveling toward the Sun propagate approximately parallel to the Sun-Earth line. In the GSE coordinate system, the x axis points toward the Sun, the z axis is toward the north ecliptic pole, and the y axis is roughly opposite Earth's orbital motion. Positively charged cosmic rays are only affected by the $y$ and $z$ components of the IMF in the GSE coordinate system, but not by the $x$ component. $B_y$ and $B_z$ cause displacement of the Sun shadow along the north-south and west-east directions, respectively, according to the Lorentz force law. However, the strength of $B_z$ is influenced not only by the IMF, but also by the geomagnetic field, which has a complex effect on the west-east displacement of the Sun shadow.[23,25] Therefore, the dominant displacement of the Sun shadow is in the north-south direction, caused by $B_y$.[24,25]

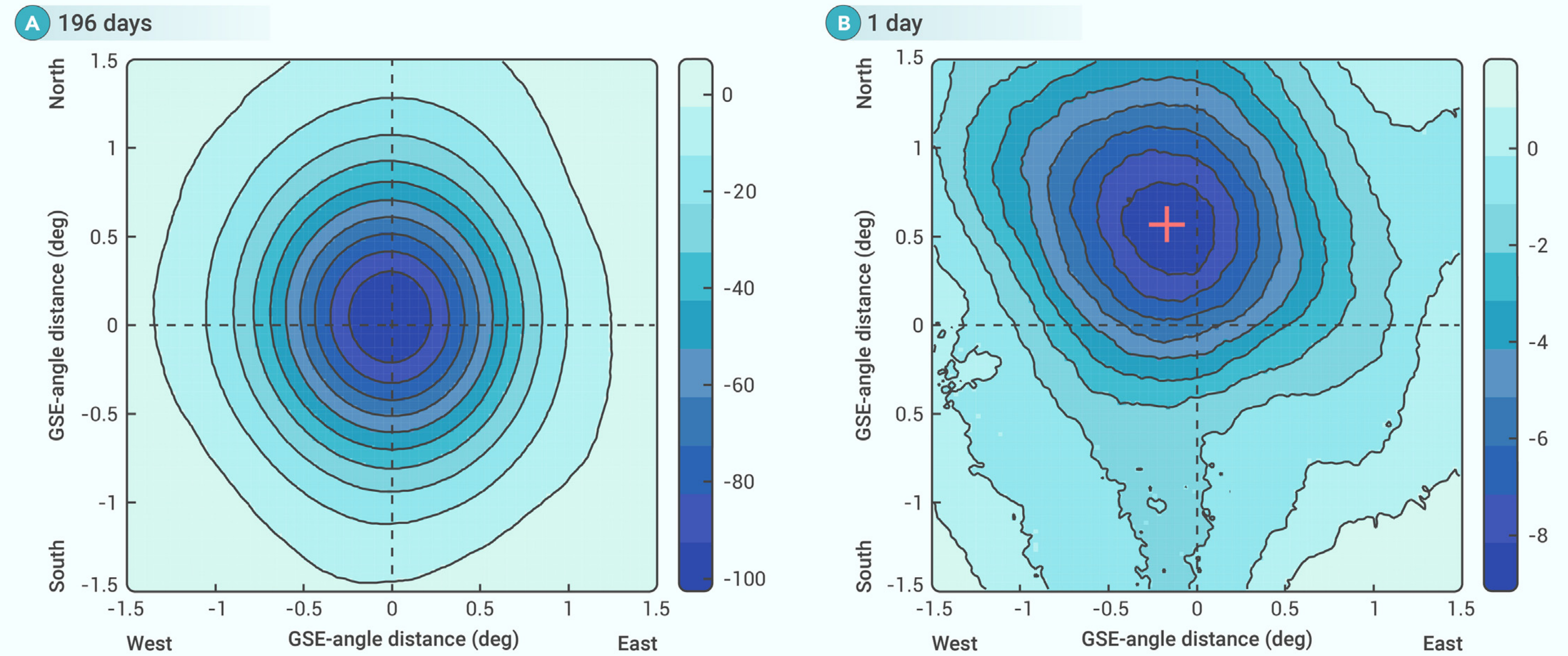

**Figure 1. Observed significance map of Sun shadows by LHAASO for two timescales** (A) Map for 196 days. The central circle of the contour map indicates a significance of $-102.5\sigma$, and the step between the contour lines is $10\sigma$. (B) Map for a single day on May 26, 2021. The central circle of the contour map indicates a significance of $-9.2\sigma$, and the step between the contour lines is $1\sigma$. The best-fit displacement of the Sun shadow shown as the red plus sign is $(-0.16^\circ \pm 0.08^\circ, 0.52^\circ \pm 0.08^\circ)$. This displacement provides a measurement of the transverse magnetic field along the Sun-Earth line.

The daily displacement of the Sun shadow along the north-south direction (denoted by $D$) and its variation can be monitored by LHAASO, as illustrated in Figure 2A, where only days with a significance exceeding $5\sigma$ are shown. The number of effective observation days for $D$ reaches 177 (90% of the total days). The variation in daily $D$ appears to be periodic in each 27-day Carrington rotation (i.e., solar rotation), with the periodicity gradually changing throughout the Carrington rotation. This allows LHAASO to directly test the specific correlation between $D$ and IMF-$B_y$, potentially enabling the use of $D$ to measure daily $B_y$ for the first time.

The $B_y$ values used in our test are observational results at $L_1$ from OMNI.[30] Figure 2B displays the daily $B_y$ values, representing the mean value of hourly $B_y$ measurements within a 24-h period. The number of efficiently observed days for $B_y$ is 181 (92% of the total observation period). As presented in Figures 2A and 2B, $D$ and $B_y$ exhibit similar trends over each Carrington rotation, with a possible time lag between them.

To determine the correlation and time lag between $D$ and $B_y$, we used the discrete correlation function (DCF) method,[31] which considers the unevenly sampled time series of $D$ and $B_y$ and their measurement errors. The time lag bin width was set to 1 day, matching the time bins of $D$ and $B_y$ in Figures 2A and 2B, and we considered time lags of up to 5 days. To achieve higher precision in the time lag determination, the cadence was set to 0.0625 days based on the hourly $B_y$ measurement and a time lag sliding technique. The DCF coefficient and its error are shown in Figure 2C. The error of the time lag was estimated using $10^3$ random time series of $D$ and $B_y$,[32] generated based on a Gaussian probability distribution. The standard deviation of the distribution of time lags between the $10^3$ random time series of $D$ and $B_y$ was taken as the error of the time lag.

For the entire dataset (labeled as "All"), we tested the correlation between 177 days of $D$ data and 181 days of $B_y$ data. $D$ is most correlated with $B_y$ at $L_1$ when $D$ precedes $B_y$ by $3.31 \pm 0.12$ days, according to the maximum DCF coefficient shown with blue markers in Figure 2C. The confidence level of the maximum DCF coefficient was estimated using the Monte Carlo method.[33,34] Specifically, we generated $10^5$ random time series $D$ by randomizing both the phase and amplitude of the Fourier transform of the observed time series $D$. The DCF was then applied to each random time series $D$ and the observed time series $B_y$. The corresponding confidence level of the maximum DCF coefficient exceeded 99.73% (corresponding to $3\sigma$) with a two-sided $p$ value of 0.0027, as indicated by the dashed line in Figure 2C.

As depicted in Figure 3B, $D$ remains correlated with $B_y$ after considering a time lag of 3.31 days. The specific correlation is fitted using the following linear formula:

$$B_y(t) = (7.6 \pm 0.6)nT/^\circ \times D(t - 3.31) + (0.2 \pm 0.1)nT. \quad \text{(Equation 1)}$$

The corresponding correlation coefficient is 0.67. The functional form of this formula is essentially the same as the change in position according to the Newton-Lorentz equation. Following Equation 1, the daily $B_y$ at $L_1$ can be estimated directly based on $D$ measured by LHAASO 3.31 days earlier. The $B_y$ estimated by $D$ reflects the effective IMF responsible for the cumulative deflection of cosmic rays along the Sun-Earth line.

Figure 3A presents a comparison of the estimated $B_y$ derived from $D$ with the $B_y$ value obtained from OMNI. We performed a $\chi^2$-test to compare these values. The $\chi^2$ value is 155.8 at 159 degrees of freedom, and the corresponding probability is 0.56. All $B_y$ estimates from $D$ and OMNI fall within 3 standard deviations, except for June 1 (3.1 standard deviations) and June 4 (3.3 standard deviations). These results indicate that daily $B_y$ and its variations can be well estimated by monitoring $D$ of the Sun shadow 3.31 days earlier.

## IMF model diagnosis from the time lag between $D$ and $B_y$

Based on the classic Parker model, the IMF in heliocentric spherical coordinates can be expressed as follows:

$$B(r,\theta,\phi) = B_r(b,\theta,\phi_0)\left(\frac{b}{r}\right)^2\left[\hat{e}_r - \frac{\omega(r-b)\sin\theta}{\nu}\hat{e}_\phi\right], \quad \text{(Equation 2)}$$

where $B_r(b,\theta,\phi_0)$ is the outermost CMF at the boundary radius $b$ and heliolongitude $\phi_0$. Beyond $b$, the IMF in the model is blown out by the radial solar wind with velocity $\nu$. When the Sun rotates with angular velocity $\omega$, the streamline of the magnetic field with azimuth $\phi_0$ at $r = b$ is given by $\frac{r}{b} - 1 - \ln\left(\frac{r}{b}\right) = \frac{\nu}{b\omega}(\phi - \phi_0)$.

To study the effect of the magnetic field predicted by the IMF model, an antiparticle retraction method was adopted to simulate the Sun shadow. The details of the simulation program are presented in the materials and methods. In the Parker model calculation, the boundary radius $b$ was set to 2.5 $R_\odot$, and $B_r(b,\theta,\phi_0)$ was extrapolated from the PFSS model[9] using a 9th-order spherical harmonic expansion.[35] Photospheric magnetograms named "mrnqs" from the Global Oscillation Network Group (GONG)[36] served as an input. The calculated IMF was modified by a factor of 5.7 according to the observed average daily $B_y$ at $L_1$ from OMNI. This scaling of magnetogram data used as the input to a solar wind model is a standard practice in the heliophysics community.[37] The Sun's rotation period was ~25.4 days and the solar wind velocity $\nu$ was obtained from the daily average value from OMNI.

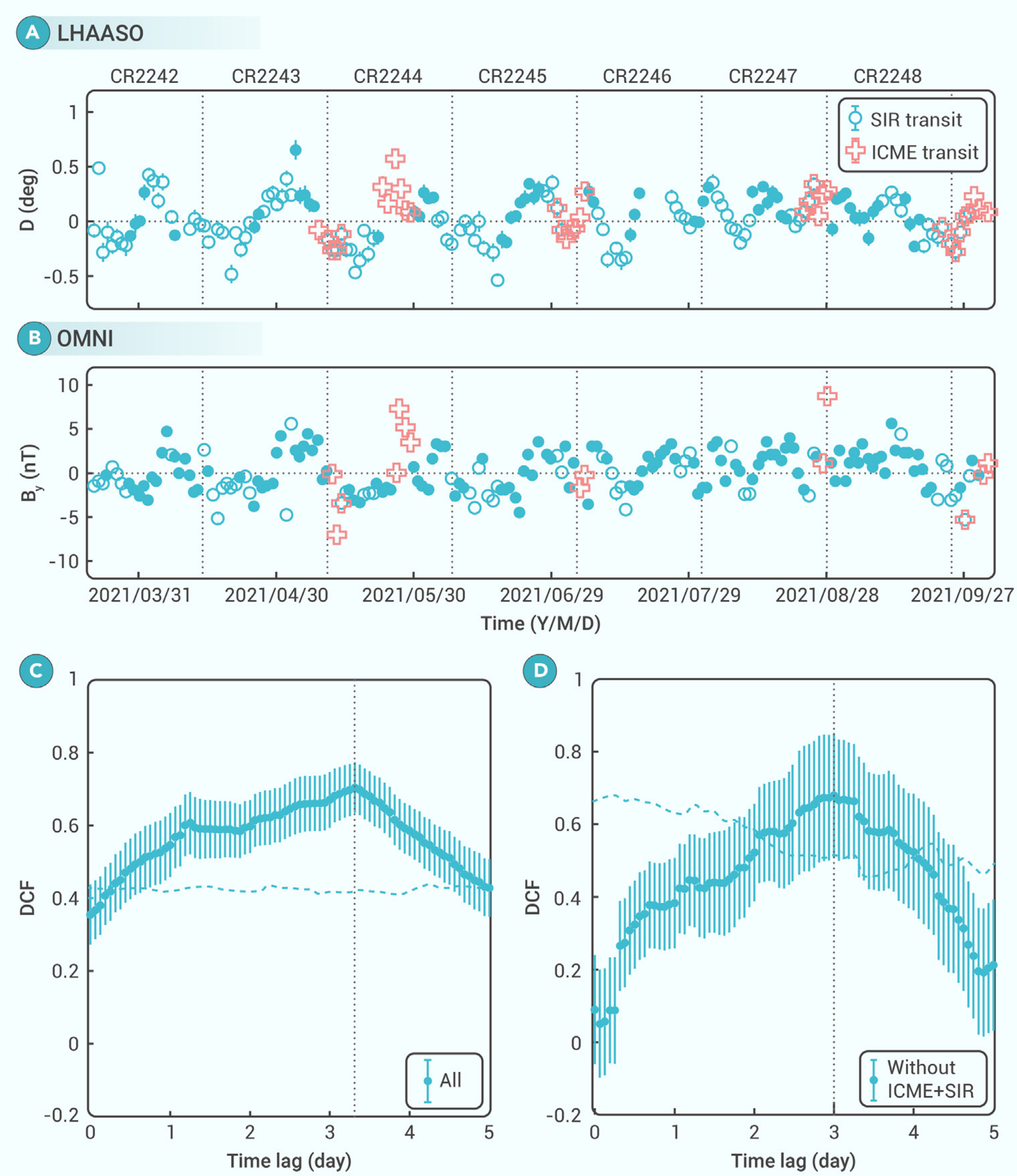

**Figure 2. Correlation between N-S displacement $D$ of the LHAASO Sun shadow and $B_y$ at $L_1$ from OMNI with 3.31-day time lag.** Daily variations of $D$ and $B_y$ are shown in (A) and (B), respectively. The red open crosses and blue open circles correspond to the periods during interplanetary coronal mass ejections (ICMEs) and stream interaction regions (SIRs) transits, respectively. The error bar shows the statistical error. The vertical dashed lines are the boundaries between Carrington rotations (CRs). The correlation coefficient of the discrete correlation function (DCF) between $D$ and $B_y$ as a function of the time lag for the "All" data sample (C) and the "without ICME and SIR" data sample (D), respectively. The dots are the discrete correlation coefficients and their error bar are included. The dashed lines are the 3$\sigma$ confidence intervals. The vertical lines show the time lag of $3.31 \pm 0.12$ days for the "All" and $3.00 \pm 0.21$ days for the "without ICME and SIR" data sample.

The IMF variations are transferred from the Sun to the Earth by the solar wind at velocity $\nu$. Cosmic rays can record $B_y$ between the Sun and Earth through $D$ of the Sun shadow. Hence, the variation in $B_y$ at $L_1$ lags behind the variation in $D$. Because $D$ represents a cumulative effect of $B_y$ that spreads from the Sun to the Earth, the specific time lag value depends on the distribution of $B_y$ along the Sun-Earth line. Therefore, the time lag between $D$ and $B_y$ at $L_1$ provides an opportunity to test the IMF models. Based on the Parker model, the average simulated time lag for the entire dataset is $2.06 \pm 0.04$ days. This simulated result reproduces the phenomenon that $D$ leads $B_y$ in the observations. However, a deviation between the simulated and observed time lags exists. This suggests a more complicated spiral structure of the IMF than that depicted by the Parker model.

One possible solution to address this deviation between the simulated and observed time lag is to add a steady, azimuthal IMF component, $B_\phi(b)$, at the CMF boundary $b$ to the Parker model, as proposed by Smith and Bieber.[38] This modified model has been used to explain the deviation of the spiral structure from the Parker model[38] and even to calculate such a deviation to interpret recent Parker Solar Probe results.[39] During the observation time of LHAASO, the additional azimuthal IMF component $B_\phi(b) \simeq -0.002 \frac{B_r(b)\nu}{\nu(b)}$, which corresponds to a "gardenhose" angle of the spiral that is ~3.6° larger than that predicted by the Parker model at $L_1$. Here, $\nu(b)$ represents the velocity at boundary $b$. Based on this modified model, the simulated time lag becomes $2.64 \pm 0.04$ days with a deviation from the observed lag. This simulated time lag is close to our measured results; however, a deviation still persists.

The magnetic field structures of CMEs can disturb the IMF to form interplanetary CMEs (ICMEs), which may lead to IMF deviations from the Parker model. In addition, solar wind interactions can disturb the IMF by generating stream interaction regions (SIRs), which may also lead to IMF deviations. To check for a possible effect of CMEs and SIRs on the deviation between the observed and simulated time lag, a data sample without ICME and SIR transits (labeled as "Without ICME+SIR") was segmented, as depicted by the blue dots in Figures 2A and 2B. We excluded ICMEs based on an updated catalog[40] and excluded SIRs based on coronal hole information[41] and an identified method,[42] as shown in Figures 2A and 2B. The ICME transit of the Sun shadow was counted from the occurrence of the CME until it stops disturbing $B_y$ at $L_1$. Similarly, the SIR transit was counted from the time when the coronal hole faces the Earth until the SIR stops disturbing $B_y$ at $L_1$. The ICME or SIR transit of $B_y$ at $L_1$ was counted from the time when the ICME or SIR starts disturbing $B_y$ at $L_1$ until it stops affecting $B_y$ at $L_1$.

For the Without ICME+SIR data sample, we tested the correlation between 59-day $D$ and 130-day $B_y$. The observed time lag between $D$ and $B_y$ at $L_1$ was $3.00 \pm 0.21$ days according to the maximum DCF coefficient, as indicated by the blue markers in Figure 2D. The corresponding confidence level of the maximum DCF coefficient exceeded 3$\sigma$ as shown by the dashed lines in Figure 2D. There was no significant difference in the observed time lag between the All and Without ICME+SIR data samples. Therefore, the deviation between the observed and simulated time lags was not due to CMEs and SIRs.

Other possible explanations for the deviation between the observed and simulated time lags include systematic changes in the IMF[43,44] and disturbances with timescales much shorter than the Carrington rotation.[45-47] A more comprehensive discussion about these possibilities is beyond the scope of this study, as it focuses on long-term average time lag.

## CONCLUSION AND OUTLOOK

Traveling almost at the speed of light, very-high-energy galactic cosmic rays take only approximately 8 min to traverse the distance between the Sun and Earth. This provides us with a useful method for exploring the magnetic field between the Sun and Earth. Based on the unprecedented monitoring of the daily cosmic-ray Sun shadow by KM2A in LHAASO, we provided daily measurements of the IMF between the Sun and Earth near the sunspot minimum period. For the first time, we made it possible for the Sun shadow to achieve an accurate measurement of the transverse magnetic field along the entire Sun-Earth line. Using this measurement, the IMF component along the GSE $y$-direction near the Earth was deduced to be $3.31 \pm 0.12$ days ahead of the measurement by spacecraft at $L_1$.

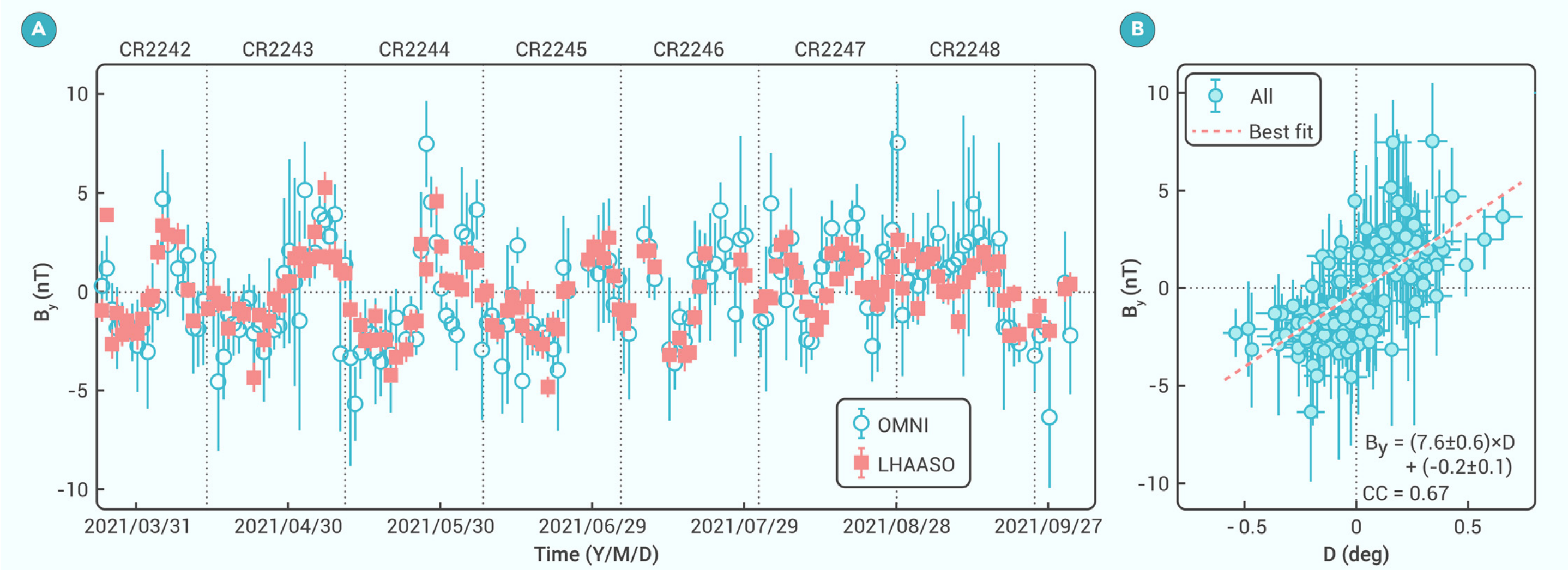

**Figure 3. Daily $B_y$ at $L_1$ as predicted by LHAASO and measured at $L_1$ 3.31 days later** (A) The red full squares and the blue open circles are for LHAASO and OMNI data, respectively. The error bar of LHAASO results is the statistical error, which includes the error from the $D$ and the fitting error from Equation 1. The error bar of OMNI results is the root mean-squared value. The vertical dashed lines are the boundaries between CRs. (B) Scatterplot of $D$ and $B_y$ at $L_1$ from OMNI after shifting for the time lag with the "All" data sample, and their correlation coefficient (CC). The error bars of $D$ indicate the statistical errors. The error bars of $B_y$ is the root mean squared value. The dashed line is the best-fit linear formula for the $B_y$ measurement by LHAASO.

With a measured $3.31 \pm 0.12$-day time lag, we also tested the classic Parker model of the IMF. We found that the measured lag significantly deviates from the model's prediction of $2.06 \pm 0.04$ days, indicating the need for refinements to the IMF model. This measurement not only provides a new method for monitoring the IMF and diagnosing existing models but also drives the development of more accurate IMF models.[4,35,38,48] This advancement will contribute to improved understanding of the propagation and forecasting of solar activity events, which can be helpful for research on space weather effects that impact human activities.

Moreover, the observed time lag implies that the Sun shadow has the potential for predicting the IMF arrival on Earth. Future research will explore the possibility of Sun shadow forecasting of the IMF component along the z axis (north-south direction) when the geomagnetic field effect is excluded. Even the possibility of forecasting a specific solar activity event, such as CMEs and SIRs, holds promise for further increasing space weather forecasting capabilities.

## MATERIALS AND METHODS

### KM2A experiment

The KM2A, a component of LHAASO, consists of 5216 EDs with 15 m spacing and 1,188 muon detectors with 30 m spacing, covering an area of 1.3 km$^2$. EDs are designed to detect showers from cosmic-ray ions and gamma rays[49] with determining their directions and energies. The 3/4 array has been operating since December 1, 2020, and the full array since July 19, 2021. During this period, the pointing accuracy, angular resolution, and energy of the cosmic rays measured by the array were stable, based on observations of the cosmic-ray Moon shadow.[29]

### Signal and displacement analysis of Sun shadow

The Sun shadow was analyzed on a sky map with $0.025° \times 0.025°$ grid spacing along the longitude and latitude in the GSE coordinate system. The background was estimated using the equal zenith angle method.[50] Specifically, there was one on-source window centered around the Sun and 20 off-source windows aligned at the same zenith angle to cover all azimuthal angles. In each grid cell of the on-source window, the background events were estimated according to the number of cosmic rays at the same grid points in the 20 off-source windows. Combined with the data selection described in the results and discussion, we collected 1.1 billion cosmic-ray events in on- and off-source windows in the data sample. The energy of these events was obtained from the energy distribution of the simulated primary cosmic rays. The median energy was $\sim$40 TeV, with a 68% interval between 22 and 107 TeV. The angular resolution containing 68% of the events was 0.5°.

The signal events were extracted using a smoothing procedure with Gaussian weighting, and the corresponding significance was estimated using the Li and Ma formula (see Equation 2.5 in Nan and Chen[51]). The significance of the Sun shadow varies with the day, with an average daily significance reaching 8.4$\sigma$, ranging from 5$\sigma$ to 13.8$\sigma$. The displacement of the Sun shadow was estimated by a likelihood ratio test between the one-source model and background-only model,[52] with the displacement of the Sun shadow along the west-east and north-south directions as free parameters.

### Monte Carlo simulation of Sun shadow

For the Monte Carlo simulations of KM2A detection of cosmic-ray showers, the cosmic-ray primary chemical composition and energy spectrum were specified according to the model of Gaisser et al.[53] The cascade processes induced by the cosmic-ray interactions within the atmosphere and the response of the detector were simulated by the CORSIKA[54] and G4KM2A[52,55] codes, respectively. Then, the primary cosmic-ray properties were reconstructed and selected using the same methods and conditions as in the basic observation[51] and in this work.

During the Sun shadow simulation, the Sun was tracked in real time. Above the atmosphere, an average of $1.27 \times 10^7$ particles with opposite charge to the primary cosmic rays were isotropically thrown back around the Sun's direction within a window of $10° \times 10°$. The simulation included magnetic fields along the Sun-Earth line and was performed on each day of data collection. The paths of the particles along the Sun-Earth line were tracked according to the changes in momentum and position following the Newton-Lorentz equation. In addition to the IMF, the CMF and geomagnetic field were considered in the Sun shadow simulation (the CMF has been described previously). The geomagnetic field calculation followed the International Geomagnetic Reference Field-13 (IGRF-13),[56] in which 13 (2) orders of spherical harmonic expansion were used for fields below (above) 600 km from the Earth's surface. When a particle hits the Sun, it is counted as a Sun shadow signal coming from opposite to the initial throwing direction. Finally, the initial throwing directions were smeared using the KM2A angular resolution.[18,57]

## ACKNOWLEDGMENTS

We would like to thank all staff members who worked at the LHAASO site, situated more than 4,400 m above sea level, for their year-round dedication to maintaining the detector and ensuring smooth operation of the water recycling system, electricity power supply, and other experimental components. We are grateful to the Chengdu Management Committee of Tianfu New Area for their constant financial support for research using LHAASO data. We appreciate the computing and data service support provided by the National High Energy Physics Data Center for the data analysis presented in this paper. We also thank the OMNI and GONG groups for providing the data, SSW and IAGA for providing the codes, and Shen Zhenning for his guidance on using Smith and Bieber's modified model. This research was supported by the following grants: The National Key R&D

Program of China (no. 2018YFA0404201), National Natural Science Foundation of China (nos. 12393851, 12393854, 12175121, 12205314, 12105301, 12305120, 12261160362, 12105294, U1931201, and 12375107), China Postdoctoral Science Foundation (no. 2022M723150), National Science and Technology Development Agency of Thailand, and National Research Council of Thailand under the High-Potential Research Team Grant Program (N42A650868).

## AUTHOR CONTRIBUTIONS

Y.N., S.C., and C.F. led the writing of the text for the data analysis and interpretation. S.C. initiated this work. Y.N. analyzed the KM2A data under the guidance of S.C. and C.F., while J.X. crosschecked the results. C.J. and Y.Y. offered expertise on solar magnetic models. Z.C. acted as spokesperson for the LHAASO Collaboration and principal investigator of the LHAASO project. All other authors participated in data analysis (including detector calibration, data processing, event reconstruction, data quality checks, and various simulations) and provided comments on the manuscript.

## DECLARATION OF INTERESTS

The authors declare no competing interests.

## SUPPLEMENTAL INFORMATION

It can be found online at https://doi.org/10.1016/j.xinn.2024.100695.

## LEAD CONTACT WEBSITE

Yuncheng Nan: https://orcid.org/0000-0002-7363-0252
Songzhan Chen: https://orcid.org/0000-0003-0703-1275
Cunfeng Feng: https://orcid.org/0000-0001-9138-3200

## Author List

Zhen Cao[1,2,3],F. Aharonian[4,5],Axikegu[6],Y.X. Bai[1,3],Y.W. Bao[7],D. Bastieri[8],X.J. Bi[1,2,3],Y.J. Bi[1,3],W. Bian[9],A.V. Bukevich[10],Q. Cao[11],W.Y. Cao[12],Zhe Cao[13,12],J. Chang[14],J.F. Chang[1,3,13],A.M. Chen[9],E.S. Chen[1,2,3],H.X. Chen[15],Liang Chen[16],Lin Chen[6],Long Chen[6],M.J. Chen[1,3],M.L. Chen[1,3,13],Q.H. Chen[6],S. Chen[17],S.H. Chen[1,2,3],S.Z. Chen[1,3],T.L. Chen[18],Y. Chen[7],N. Cheng[1,3],Y.D. Cheng[1,2,3],M.Y. Cui[14],S.W. Cui[11],X.H. Cui[19],Y.D. Cui[20],B.Z. Dai[17],H.L. Dai[1,3,13],Z.G. Dai[12],Danzengluobu[18],X.Q. Dong[1,2,3],K.K. Duan[14],J.H. Fan[8],Y.Z. Fan[14],J. Fang[17],J.H. Fang[15],K. Fang[1,3],C.F. Feng[21],H. Feng[1],L. Feng[14],S.H. Feng[1,3],X.T. Feng[21],Y. Feng[15],Y.L. Feng[18],S. Gabici[22],B. Gao[1,3],C.D. Gao[21],Q. Gao[18],W. Gao[1,3],W.K. Gao[1,2,3],M.M. Ge[17],L.S. Geng[1,3],G. Giacinti[9],G.H. Gong[23],Q.B. Gou[1,3],M.H. Gu[1,3,13],F.L. Guo[16],X.L. Guo[6],Y.Q. Guo[1,3],Y.Y. Guo[14],Y.A. Han[24],M. Hasan[1,2,3],H.H. He[1,2,3],H.N. He[14],J.Y. He[14],Y. He[6],Y.K. Hor[20],B.W. Hou[1,2,3],C. Hou[1,3],X. Hou[25],H.B. Hu[1,2,3],Q. Hu[12,14],S.C. Hu[1,3,26],D.H. Huang[6],T.Q. Huang[1,3],W.J. Huang[20],X.T. Huang[21],X.Y. Huang[14],Y. Huang[1,2,3],X.L. Ji[1,3,13],H.Y. Jia[6],K. Jia[21],K. Jiang[13,12],X.W. Jiang[1,3],Z.J. Jiang[17],M. Jin[6],M.M. Kang[27],I. Karpikov[10],D. Kuleshov[10],K. Kurinov[10],B.B. Li[11],C.M. Li[7],Cheng Li[13,12],Cong Li[1,3],D. Li[1,2,3],F. Li[1,3,13],H.B. Li[1,3],H.C. Li[1,3],Jian Li[12],Jie Li[1,3,13],K. Li[1,3],S.D. Li[16,2],W.L. Li[21],W.L. Li[9],X.R. Li[1,3],Xin Li[13,12],Y.Z. Li[1,2,3],Zhe Li[1,3],Zhuo Li[28],E.W. Liang[29],Y.F. Liang[29],S.J. Lin[20],B. Liu[12],C. Liu[1,3],D. Liu[21],D.B. Liu[9],H. Liu[6],H.D. Liu[24],J. Liu[1,3],J.L. Liu[1,3],M.Y. Liu[18],R.Y. Liu[7],S.M. Liu[6],W. Liu[1,3],Y. Liu[8],Y.N. Liu[23],Q. Luo[20],Y. Luo[9],H.K. Lv[1,3],B.Q. Ma[28],L.L. Ma[1,3],X.H. Ma[1,3],J.R. Mao[25],Z. Min[1,3],W. Mitthumsiri[30],H.J. Mu[24],Y.C. Nan[1,3],A. Neronov[22],L.J. Ou[8],P. Pattarakijwanich[30],Z.Y. Pei[8],J.C. Qi[1,2,3],M.Y. Qi[1,3],B.Q. Qiao[1,3],J.J. Qin[12],A. Raza[1,2,3],D. Ruffolo[30],A. S'aiz[30],M. Saeed[1,2,3],D. Semikoz[22],L. Shao[11],O. Shchegolev[10,31],X.D. Sheng[1,3],F.W. Shu[32],H.C. Song[28],Yu.V. Stenkin[10,31],V. Stepanov[10],Y. Su[14],D.X. Sun[12,14],Q.N. Sun[6],X.N. Sun[29],Z.B. Sun[33],J. Takata[34],P.H.T. Tam[20],Q.W. Tang[32],R. Tang[9],Z.B. Tang[13,12],W.W. Tian[2,19],C. Wang[33],C.B. Wang[6],G.W. Wang[12],H.G. Wang[8],H.H. Wang[20],J.C. Wang[25],Kai Wang[7],Kai Wang[34],L.P. Wang[1,2,3],L.Y. Wang[1,3],P.H. Wang[6],R. Wang[21],W. Wang[20],X.G. Wang[29],X.Y. Wang[7],Y. Wang[6],Y.D. Wang[1,3],Y.J. Wang[1,3],Z.H. Wang[27],Z.X. Wang[17],Zhen Wang[9],Zheng Wang[1,3,13],D.M. Wei[14],J.J. Wei[14],Y.J. Wei[1,2,3],T. Wen[17],C.Y. Wu[1,3],H.R. Wu[1,3],Q.W. Wu[34],S. Wu[1,3],X.F. Wu[14],Y.S. Wu[12],S.Q. Xi[1,3],J. Xia[12,14],G.M. Xiang[16,2],D.X. Xiao[11],G. Xiao[1,3],Y.L. Xin[6],Y. Xing[16],D.R. Xiong[25],Z. Xiong[1,2,3],D.L. Xu[9],R.F. Xu[1,2,3],R.X. Xu[28],W.L. Xu[27],L. Xue[21],D.H. Yan[17],J.Z. Yan[14],T. Yan[1,3],C.W. Yang[27],C.Y. Yang[25],F. Yang[11],F.F. Yang[1,3,13],L.L. Yang[20],M.J. Yang[1,3],R.Z. Yang[12],W.X. Yang[8],Y.H. Yao[1,3],Z.G. Yao[1,3],L.Q. Yin[1,3],N. Yin[21],X.H. You[1,3],Z.Y. You[1,3],Y.H. Yu[12],Q. Yuan[14],H. Yue[1,2,3],H.D. Zeng[14],T.X. Zeng[1,3,13],W. Zeng[17],M. Zha[1,3],B.B. Zhang[7],F. Zhang[6],H. Zhang[9],H.M. Zhang[7],H.Y. Zhang[1,3],J.L. Zhang[19],Li Zhang[17],P.F. Zhang[17],P.P. Zhang[12,14],R. Zhang[12,14],S.B. Zhang[2,19],S.R. Zhang[11],S.S. Zhang[1,3],X. Zhang[7],X.P. Zhang[1,3],Y.F. Zhang[6],Yi Zhang[1,14],Yong Zhang[1,3],B. Zhao[6],J. Zhao[1,3],L. Zhao[13,12],L.Z. Zhao[11],S.P. Zhao[14],X.H. Zhao[25],F. Zheng[33],W.J. Zhong[7],B. Zhou[1,3],H. Zhou[9],J.N. Zhou[16],M. Zhou[32],P. Zhou[7],R. Zhou[27],X.X. Zhou[1,2,3],X.X. Zhou[6],B.Y. Zhu[12,14],C.G. Zhu[21],F.R. Zhu[6],H. Zhu[19],K.J. Zhu[1,2,3,13],Y.C. Zou[34],X. Zuo[1,3],

(The LHAASO Collaboration)

C.W. Jiang[35] and Y. Yang[33].

[1] Key Laboratory of Particle Astrophysics & Experimental Physics Division & Computing Center, Institute of High Energy Physics, Chinese Academy of Sciences, Beijing 100049 , China

[2] University of Chinese Academy of Sciences, Beijing 100049 , China

[3] TIANFU Cosmic Ray Research Center, Chengdu, Sichuan, China

[4] Dublin Institute for Advanced Studies, 31 Fitzwilliam Place, 2 Dublin, Ireland

[5] Max-Planck-Institut for Nuclear Physics, P.O. Box 103980, 69029 Heidelberg, Germany

[6] School of Physical Science and Technology & School of Information Science and Technology, Southwest

Jiaotong University, Chengdu 610031, Sichuan, China
[7] School of Astronomy and Space Science, Nanjing University, Nanjing 210023, Jiangsu, China
[8] Center for Astrophysics, Guangzhou University, Guangzhou 510006, Guangdong, China
[9] Tsung-Dao Lee Institute & School of Physics and Astronomy, Shanghai Jiao Tong University, Shanghai 200240, China
[10] Institute for Nuclear Research of Russian Academy of Sciences, Moscow 117312, Russia
[11] Hebei Normal University, Shijiazhuang 050024, Hebei, China
[12] University of Science and Technology of China, Hefei 230026, Anhui, China
[13] State Key Laboratory of Particle Detection and Electronics, China
[14] Key Laboratory of Dark Matter and Space Astronomy & Key Laboratory of Radio Astronomy, Purple Mountain Observatory, Chinese Academy of Sciences, Nanjing 210023, Jiangsu, China
[15] Research Center for Astronomical Computing, Zhejiang Laboratory, Hangzhou 311121, Zhejiang, China
[16] Key Laboratory for Research in Galaxies and Cosmology, Shanghai Astronomical Observatory, Chinese Academy of Sciences, Shanghai 200030, China
[17] School of Physics and Astronomy, Yunnan University, Kunming 650091, Yunnan, China
[18] Key Laboratory of Cosmic Rays (Tibet University), Ministry of Education, Lhasa 850000, Tibet, China
[19] Key Laboratory of Radio Astronomy and Technology, National Astronomical Observatories, Chinese Academy of Sciences, Beijing 100101, China
[20] School of Physics and Astronomy (Zhuhai) & School of Physics (Guangzhou) & Sino-French Institute of Nuclear Engineering and Technology (Zhuhai), Sun Yat-sen University, Zhuhai 519000& Guangzhou 510275, Guangdong, China
[21] Institute of Frontier and Interdisciplinary Science, Shandong University, Qingdao 266237, Shandong, China
[22] APC, Universit'e Paris Cit'e, CNRS/IN2P3, CEA/IRFU, Observatoire de Paris, 119 75205 Paris, France
[23] Department of Engineering Physics, Tsinghua University, Beijing 100084, China
[24] School of Physics and Microelectronics, Zhengzhou University, Zhengzhou 450001, Henan, China
[25] Yunnan Observatories, Chinese Academy of Sciences, Kunming 650216, Yunnan, China
[26] China Center of Advanced Science and Technology, Beijing 100190, China
[27] College of Physics, Sichuan University, Chengdu 610065, Sichuan, China
[28] School of Physics, Peking University, Beijing 100871, China
[29] Guangxi Key Laboratory for Relativistic Astrophysics, School of Physical Science and Technology, Guangxi University, Nanning 530004, Guangxi, China
[30] Department of Physics, Faculty of Science, Mahidol University, Bangkok 10400, Thailand
[31] Moscow Institute of Physics and Technology, Moscow 141700, Russia
[32] Center for Relativistic Astrophysics and High Energy Physics, School of Physics and Materials Science & Institute of Space Science and Technology, Nanchang University, Nanchang 330031, Jiangxi, China
[33] National Space Science Center, Chinese Academy of Sciences, Beijing 100190, China
[34] School of Physics, Huazhong University of Science and Technology, Wuhan 430074, Hubei, China
[35] Institute of Space Science and Applied Technology, Harbin Institute of Technology, Shenzhen 518055, China